\documentclass[preprint,showpacs,prd]{revtex4}
\usepackage{epsfig}

\begin{document}

%\large

\title{\boldmath Search for the Pentaquark State in 
$\psi(2S)$ and $J/\psi$ Decays to $K^0_SpK^-\bar n$ and $K^0_S\bar p K^+n$}
\author{\small J.~Z.~Bai$^1$,        Y.~Ban$^{10}$,         J.~G.~Bian$^1$,
X.~Cai$^{1}$,         J.~F.~Chang$^1$,       H.~F.~Chen$^{16}$,
H.~S.~Chen$^1$,       H.~X.~Chen$^{1}$,      J.~Chen$^{1}$,
J.~C.~Chen$^1$
% ...
\vspace{0.1cm}
\\(BES Collaboration)\\
\vspace{0.2cm}
$^1$ Institute of High Energy Physics, Beijing 100039, People's
Republic of     China\\
$^2$ China Center of Advanced Science and Technology, Beijing 100080,
     People's Republic of China\\
$^3$ Guangxi Normal University, Guilin 541004, People's Republic of
China\\
$^4$ Henan Normal University, Xinxiang 453002, People's Republic of
China\\
$^5$ Huazhong Normal University, Wuhan 430079, People's Republic of
China\\
$^6$ Hunan University, Changsha 410082, People's Republic of China\\
$^7$ Liaoning University, Shenyang 110036, People's Republic of
China\\
$^{8}$ Nanjing Normal University, Nanjing 210097, People's Republic of
China\\
$^{9}$ Nankai University, Tianjin 300071, People's Republic of China\\
$^{10}$ Peking University, Beijing 100871, People's Republic of
China\\
$^{11}$ Shandong University, Jinan 250100, People's Republic of
China\\
$^{12}$ Shanghai Jiaotong University, Shanghai 200030,
        People's Republic of China\\
$^{13}$ Sichuan University, Chengdu 610064,
        People's Republic of China\\
$^{14}$ Tsinghua University, Beijing 100084,
        People's Republic of China\\
$^{15}$ University of Hawaii, Honolulu, Hawaii 96822\\
$^{16}$ University of Science and Technology of China, Hefei 230026,
        People's Republic of China\\
$^{17}$ Wuhan University, Wuhan 430072, People's Republic of China\\
$^{18}$ Zhejiang University, Hangzhou 310028, People's Republic of
China\\
\vspace{0.4cm}
$^{\ast}$ Visiting professor to University of Michigan, Ann Arbor, MI
48109 USA
}

\vspace*{0.4cm}
\date{\today}

\begin{abstract}
Results are presented on $\psi(2S)$ and $J/\psi$ hadronic decays to
$K^0_SpK^-\bar n$ and $K^0_S\bar p K^+n$ final states from data
samples of 14 million $\psi(2S)$ and 58 million $J/\psi$ events accumulated
at the BES\,II detector. No $\Theta(1540)$ signal, the
pentaquark candidate, is observed, and upper limits for ${\cal
B}(\psi(2S)\to\Theta\bar\Theta\to K^0_S p K^-\bar n + K^0_S \bar p K^+
n) < 0.84\times 10^{-5}$ and ${\cal B}(J/\psi\to\Theta\bar\Theta\to K^0_S
p K^-\bar n + K^0_S \bar p K^+ n) < 1.1\times 10^{-5}$ at the 90\%
confidence level are set. For single $\Theta(1540)$ production, the
upper limits determined by our analysis are also on the order of
$10^{-5}$ in both $\psi(2S)$ and $J/\psi$ decays.
\end{abstract}

%\draft
%\vspace{10mm}
%\input{a}
%\author{D.V. Bugg$^{21}$\\
%{\small\it $^{21}$Queen Mary, London E1\,4NS, UK\\}}
%\maketitle

%%%%\vspace{0.4cm}

%{\flushleft PACS: {13.25.Gv, 14.20.Jn}}% PACS, the Physics and Astronomy}
\maketitle

%\section{\label{sec:level1}First-level heading:\protect\\ The line
%break was forced \lowercase{via} \textbackslash\textbackslash}

%\vspace*{5pt}
\section{Introduction}

Recently the LEPS Collaboration at Spring-8 reported the $4.6~\sigma$
discovery of a new $S = +1$ state, the $\Theta(1540)$, with a mass of
$1.54\pm0.01$ GeV$/c^2$ and a width of less than 25 MeV$/c^2$, close
to $NK$ threshold in the reaction $\gamma ^{12}C\to K^+K^-X$
\cite{LEPS}.  Subsequently, the DIANA Collaboration at ITEP, CLAS at
Jefferson Lab, and SAPHIR at ELSA claimed this narrow state, the
candidate for a pentaquark state ($uudd\bar s$), in $n + K^+$ or $p +
K^0$ decay configurations - all published in 2003 [2-4].

The work of LEPS was motivated, in part, by the work by Diakonov,
Petrov and Polyakov who studied anti-decuplet baryons using the chiral
soliton model \cite{DIA}. In their model, the anti-decuplet was
anchored to the mass of the $P_{11}(1710)$ nucleon resonance, giving
the pentaquark state $\Theta^+$ (spin 1/2, isospin 0, and $S = +1$) a
mass of $\sim 1530$ MeV$/c^2$ and a total width of less than 15
MeV$/c^2$. In particular, it is predicted to be an isoscalar.

There are many other theoretical works to try to explain the
properties of the $\Theta(1540)$ with various quark models [6-9] or
alternative approaches \cite{J}.  Isospins 0 and 1 are
both possible; isospin 1 would lead to three charge states $\Theta^0$,
$\Theta^+$, $\Theta^{++}$. A reanalysis \cite{KN} of older
experimental data on the $K^+$-nucleon elastic scattering put a more
stringent constraint on the width to be $\Gamma_{\Theta} < 1$ MeV.
Since the mass of the $\Theta$ is larger than the sum of the masses of
the nucleon and kaon, it is not easy to understand why its width
should be so narrow, unless it has very special quantum
numbers. Capstick, Page and Roberts point out that the $\Theta^+$
could be a member of an isospin quintet with charges from $-1$ to $+3$
where the $Q = +3$ state has a $(uuuu\bar s)$ quark-model
configuration. Decays of an isotensor $\Theta^+$ into $pK^0$ and
$pK^+$ are isospin violating; hence an isotensor $\Theta^+$ is
expected to be narrow \cite{PRP}. The analyses of CLAS and SAPHIR
support that the $\Theta^+$ is isoscalar, but the statistics of
present experiments are limited.

There are experimental questions concerning the $\Theta(1540)$.  For
the four experiments [1-4], the exact shape of the background is very
difficult to estimate since the $\Theta^+$ is close to $NK$ threshold,
and strong cuts have been applied to the event samples in all
cases. The calculation of Dzierba {\it et al.} \cite{D} shows that
kinematic reflections of meson resonances could well account for the
enhancement observed in the $K^+n$ effective mass distribution at the
mass of the purported $\Theta^+$. They suggest that further experimental
studies will be required with higher statistics, including varying the
incident beam momentum and establishing the spin and parity, before
claiming solid evidence for a $S = +1$ baryon resonance.  It is
important to understand the properties of the $\Theta$ through
systematic studies by different experiments.  Compared with the above
experiments, the data accumulated at the $e^+e^-$ collision experiment
BES are relatively clean and have less background; therefore it is
meaningful to investigate the pentaquark state $\Theta$ with the
hadronic decays of the charmonium states $\psi(2S)$ and $J/\psi$.

In this work, we search for the pentaquark state $\Theta(1540)$ in
$\psi(2S)$ and $J/\psi$ decays to $K^0_SpK^-\bar n$ and $K^0_S\bar p
K^+n$ final states using samples of 14 million $\psi(2S)$ and 58
million $J/\psi$ events taken with the upgraded Beijing Spectrometer
(BES\,II) located at the Beijing Electron Positron Collider
(BEPC). These processes could contain $\Theta$ decays to
$K^0_Sp,~K^+n~(uudd\bar s)$ and $\bar\Theta$ decays to $K^0_S\bar
p,~K^-n~(\bar u\bar u\bar d\bar d s)$.

%\vspace*{20pt}
\section{Bes detector}
BES\,II is a large
solid-angle magnetic spectrometer that is described in detail in Ref.
\cite{BESII}. Charged particle momenta are determined with a
resolution of $\sigma_p/p = 1.78\%\sqrt{1+p^2(\mbox{GeV}^2)}$ in a
40-layer cylindrical drift chamber. Particle identification is
accomplished by specific ionization ($dE/dx$) measurements in the
drift chamber and time-of-flight (TOF) measurements in a barrel-like
array of 48 scintillation counters. The $dE/dx$ resolution is
$\sigma_{dE/dx} = 8.0\%$; the TOF resolution is $\sigma_{TOF} = 180$
ps for Bhabha events. Outside of the time-of-flight counters is a
12-radiation-length barrel shower counter (BSC) comprised of gas
proportional tubes interleaved with lead sheets. The BSC measures the
energies of photons with a resolution of
$\sigma_E/E\simeq 21\%/\sqrt{E(\mbox{GeV})}$. Outside the solenoidal
coil, which provides a 0.4 Tesla magnetic field over the tracking
volume, is an iron flux return that is instrumented with three double layers of counters that are
used to identify muons.

In this analysis, a
 GEANT3 based Monte Carlo simulation package (SIMBES) with detailed
   consideration of detector performance (such as dead
   electronic channels) is used.
    The consistency between data and Monte Carlo has been
 checked in
   many high purity physics channels, and the agreement is quite
 reasonable.

%\section{Bes detector}

\section{Event selection}
In $\psi(2S)$ or $J/\psi$ decays to $K^0_SpK^-\bar n$ and $K^0_S\bar p
K^+n$, the anti-neutron and neutron are not detected.  The first step
in the analysis is to select in four prong events the $\pi^+\pi^-$
pair, which composes the $K^0_S$. Next the missing mass is calculated
according to energy-momentum conservation. The events with missing
mass close to the $\bar n (n)$'s mass are selected.  We use the same
criteria and treatment for both $\psi(2S)$ and $J/\psi$ data.

The first level of event selection requires two positively and two
negatively charged tracks with less than ten neutral tracks. The
charged particles are each required to lie within the acceptance of
the detector and to have a good helix fit.

The $K^0_S$ meson in the event is identified through the decay
$K^0_S\to\pi^+\pi^-$. For the selection of the $K^0_S$,  both the $\pi^+$ and
$\pi^-$ are required to be consistent with being pions, namely,
$Prob_{pid}(\pi) > 0.01$. The definition of $Prob_{pid}$ is
\begin {eqnarray*} 
\chi^2_{com} =
\chi^2_{TOF}+\chi^2_{dE/dx}\\
Prob_{pid} = Prob(\chi^2_{com}, 2),
\end {eqnarray*}
where $\chi^2_{TOF}$ and $\chi^2_{dE/dx}$ are determined from the
measured and expected TOF times and ionization information for the
particle hypothesis of interest.
If only TOF or $dE/dx$ information is available, $Prob_{pid} =
Prob(\chi^2,1)$.
The four charged tracks can be grouped into a maximum of four
possible $\pi^+\pi^-$ combinations. The combination with the invariant
mass closest to the mass of $K^0_S$ is chosen as the $K^0_S$
candidate, and the intersection point of the two tracks is regarded as
the secondary vertex.  Candidate events are required to satisfy
$|M_{\pi^+\pi^-}-M_{K^0_S}| < 15$ MeV$/c^2$, where $M_{\pi^+\pi^-}$ is
calculated at the $K^0_S$ decay vertex.
For the proton (or anti-proton) and kaon,
we require
$Prob_{pid}(p) > 0.01$ and $Prob_{pid}(K) > 0.01$, respectively.

Candidate events are kinematically fitted (one constraint fit) under
the assumption of a missing $\bar n(n)$ to obtain better mass
resolution and to suppress the backgrounds. As an example, Fig. 1
shows the $\chi^2_{K^0_S p K^-\bar n}$ distribution of $\psi(2S)\to K^0_S p K^-\bar n$
candidate events after particle identification. We require
$\chi^2_{K^0_SpK^-\bar n}$ or $\chi^2_{K^0_S\bar pK^+ n} < 5$ and use
a further cut on the $K^0_S$ decay length, $L_{xy} >$~3 mm, to remove
remaining backgrounds.

\begin{figure}[htbp!]
\begin{center}
\epsfxsize=8.5cm\epsffile{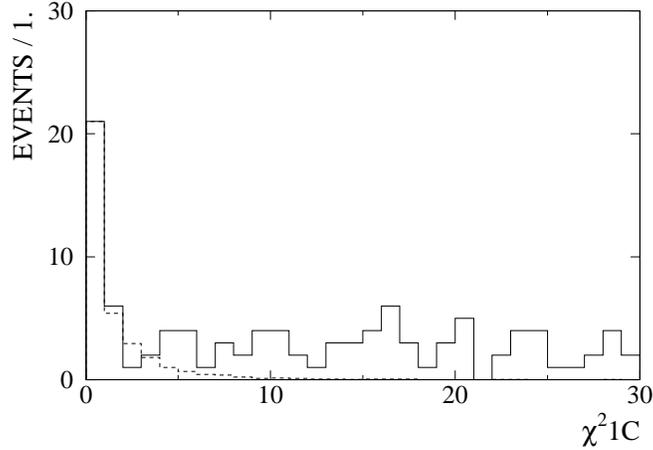}
\vspace*{5pt}
\caption{$\chi^2$ distribution of $\psi(2S)\to K^0_S p K^-\bar n$ 
candidate
events after requiring particle identification. The
solid histogram is for data, and the dashed histogram for Monte
Carlo simulation (phase space), where the Monte Carlo is normalized to
the data in the first bin. }
\end{center}
\end{figure}

Fig. 2 shows the $\bar n$ missing mass distributions of data and Monte
Carlo for $\psi(2S)\to K^0_S p K^-\bar n$ after the above cuts. The
surviving events fall mostly within the range 0.9 - 1.0 GeV$/c^2$, and
we further require 0.9 GeV$/c^2$ $< m_{missing} <$ 1.0 GeV$/c^2$. The
missing mass distributions for the other modes $\psi(2S)\to K^0_S\bar
p K^+n$ and $J/\psi\to K^0_SpK^-\bar n$, $K^0_S\bar p K^+n$ are
similar to those of Fig.~2.

\section{Analysis Results}

After the above requirements, the individual mass distributions of
$\psi(2S)\to K^0_S p K^-\bar n$ and $K^0_S \bar p K^+ n$ are shown in
Fig. 3, and the scatter plot of $K^-n~(K^0_S \bar p)$ versus $K^0_S
p~(K^+ n)$ for
$\psi(2S)\to K^0_SpK^-\bar n$ + $K^0_S\bar p K^+n$ modes is shown in
Fig.~4.  No clear $\Theta$ signal is observed in Fig. 3, which
contains 19 ($K^0_S p K^-\bar n$) and 10 ($K^0_S \bar p K^+n$)
events, or Fig. 4. We determine an upper limit for ${\cal
B}(\psi(2S)\to\Theta\bar\Theta\to K^0_S p K^-\bar n + K^0_S \bar p K^+
n)$.  The signal region is shown as a square centered at $(1.540,
1.540)$ GeV$/c^2$ in Fig. 4. Zero events fall within the signal region
defined
as $\pm20$ MeV from the central value \cite{WIDTH}. We set an upper
limit of 2.30 events in the absence of background at the 90\%
confidence level (C.L.) for $N_{\Theta\bar\Theta}$ and 
\begin{figure}[htbp]
\begin{center}
\epsfxsize=6.05cm\epsffile{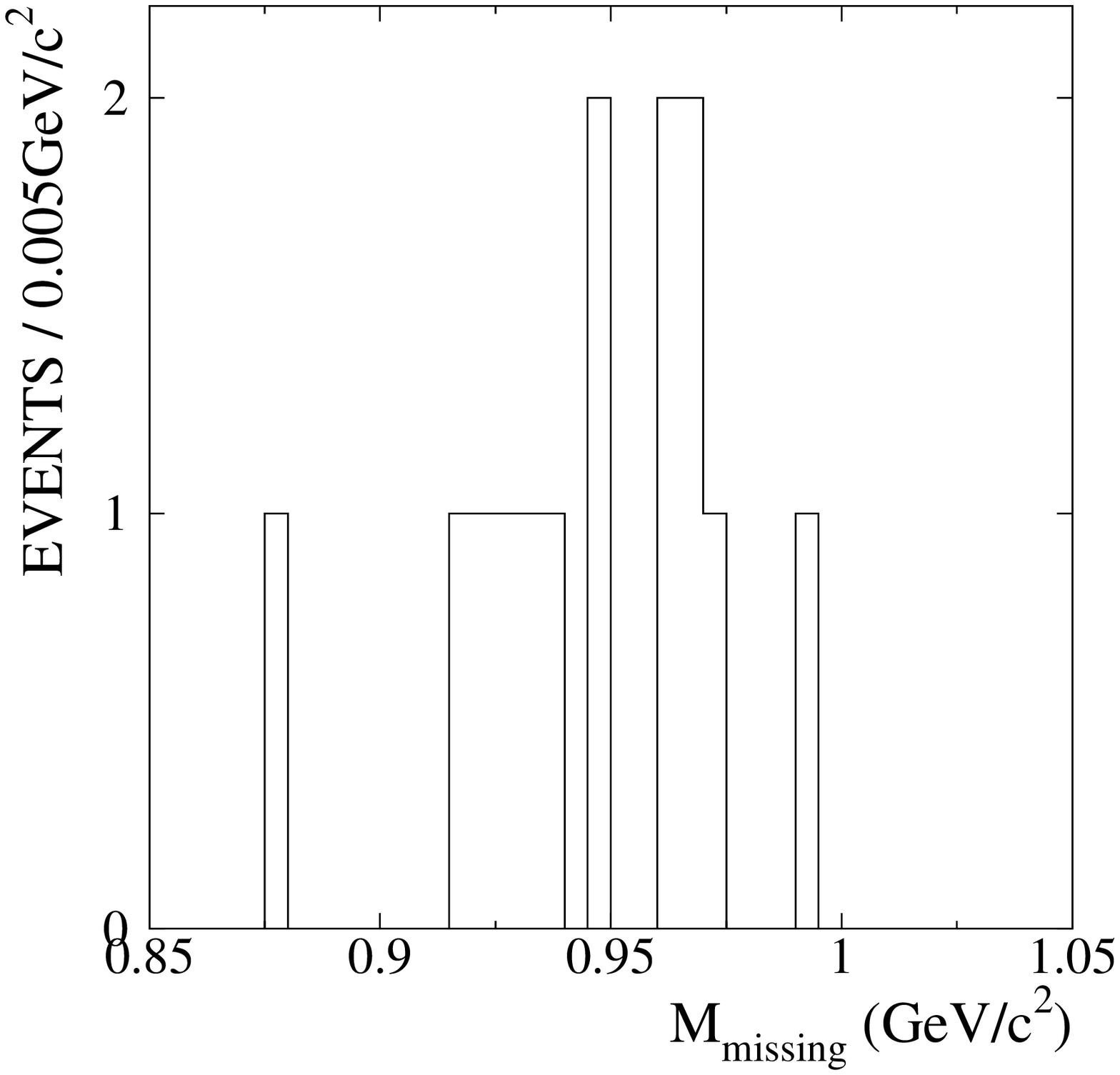}
\hspace*{0.5cm}
\epsfxsize=5.95cm\epsffile{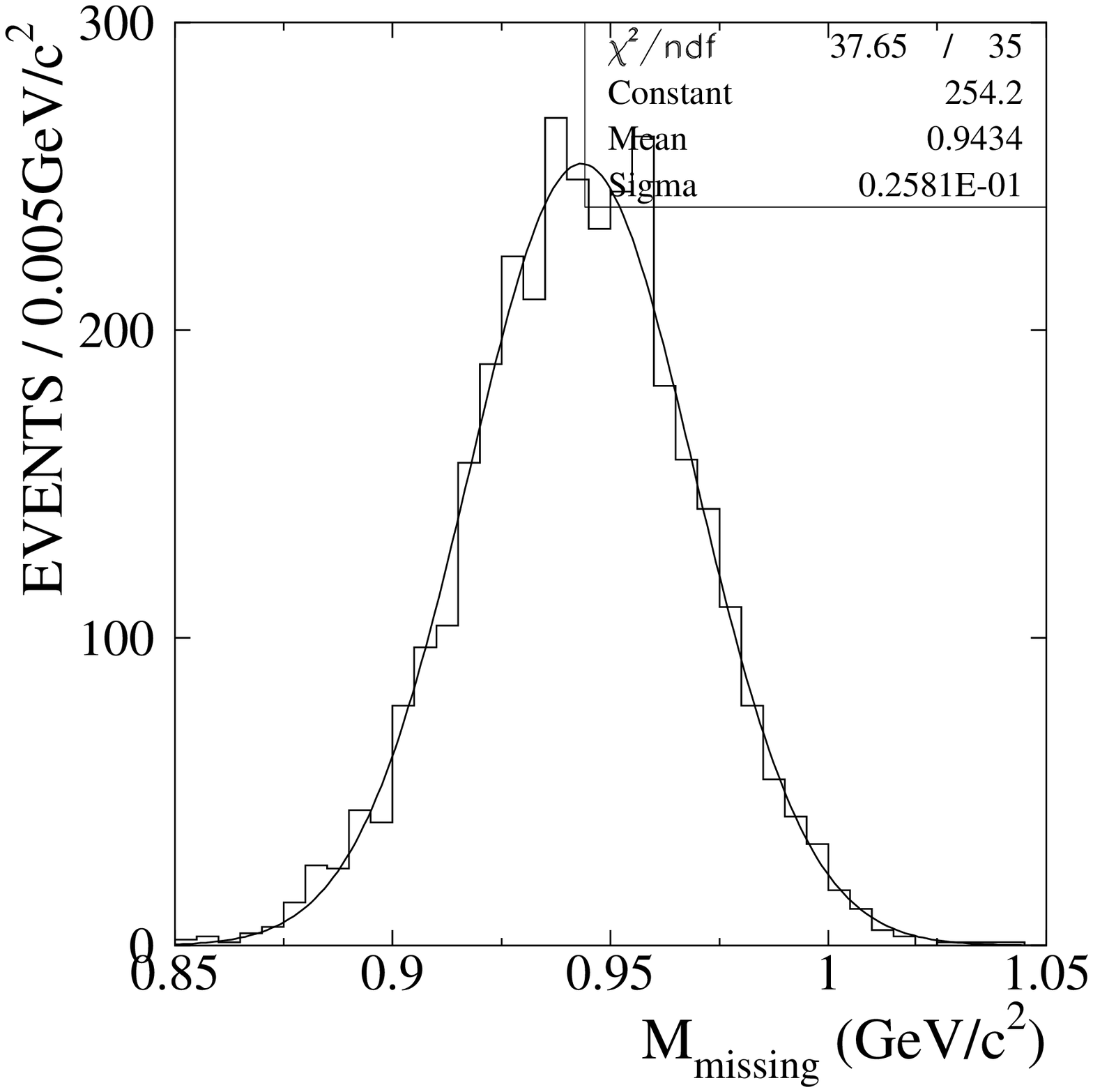}
\vspace*{5pt}
\caption{The missing mass distributions
of data (left) and Monte Carlo simulation (right) for $\psi(2S) \to
K^0_S p
K^-\bar n$.}
%\vspace*{15pt}
\end{center}
\end{figure}
$${\cal
B}(\psi(2S)\to\Theta\bar\Theta\to K^0_S p K^-\bar n + K^0_S \bar p K^+
n)$$ \vspace*{-10pt}$$< {{2.30}\over{0.686\times (2.85\pm0.08)\%\times
(14.0\times 10^{6})}}=0.84\times 10^{-5},$$\\
where 0.686 is the decay
ratio of $K^0_S$ to $\pi^+\pi^-$ \cite{PDG}; $(2.85\pm0.08)\%$ is the
detection
efficiency and the uncertainty is statistical error of the Monte Carlo
sample, and $14.0\times 10^{6}$ is the total number of
BES\,II $\psi(2S)$ events \cite{moxh}.

\begin{figure}[htbp]
\begin{center}
\epsfxsize=12.5cm\epsffile{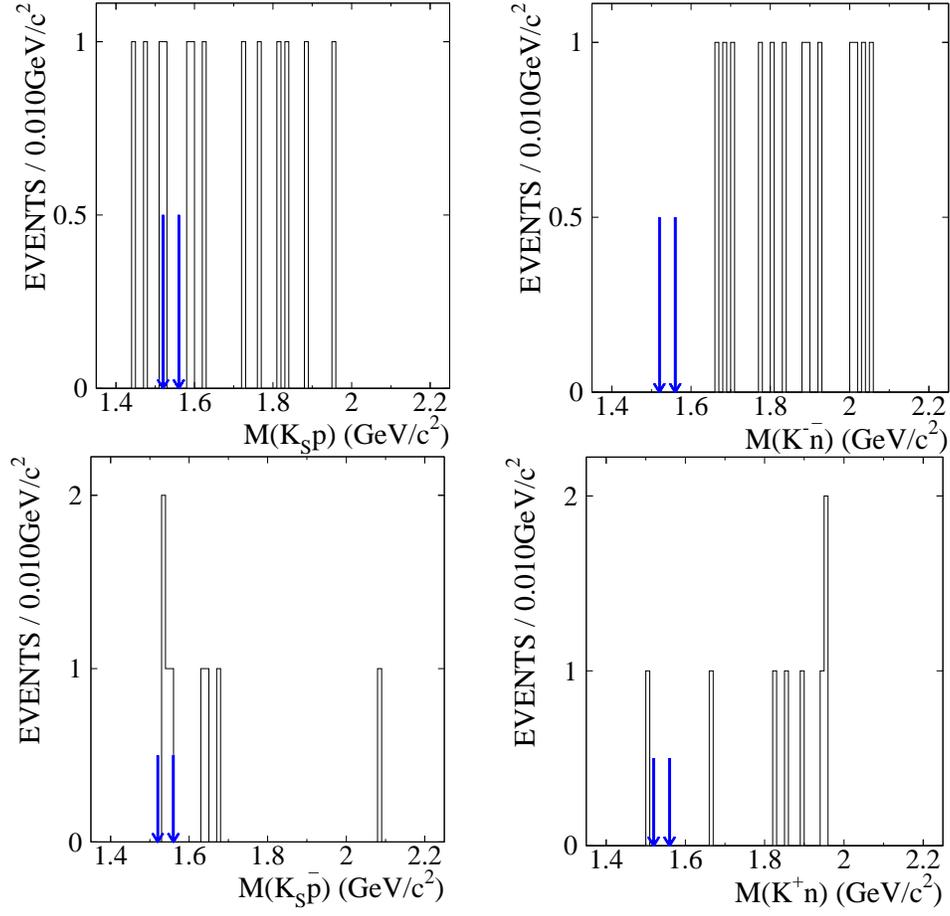}
\label{cpkp}
\vspace*{5pt}
\caption{Mass distributions of $\psi(2S){\to}K^0_S p K^-\bar
n$   
and $K^0_S \bar p K^+n$ modes.
}
\end{center}
\vspace*{-10pt}
\end{figure}

\begin{figure}[htbp]
\begin{center}
\epsfxsize=7.50cm\epsffile{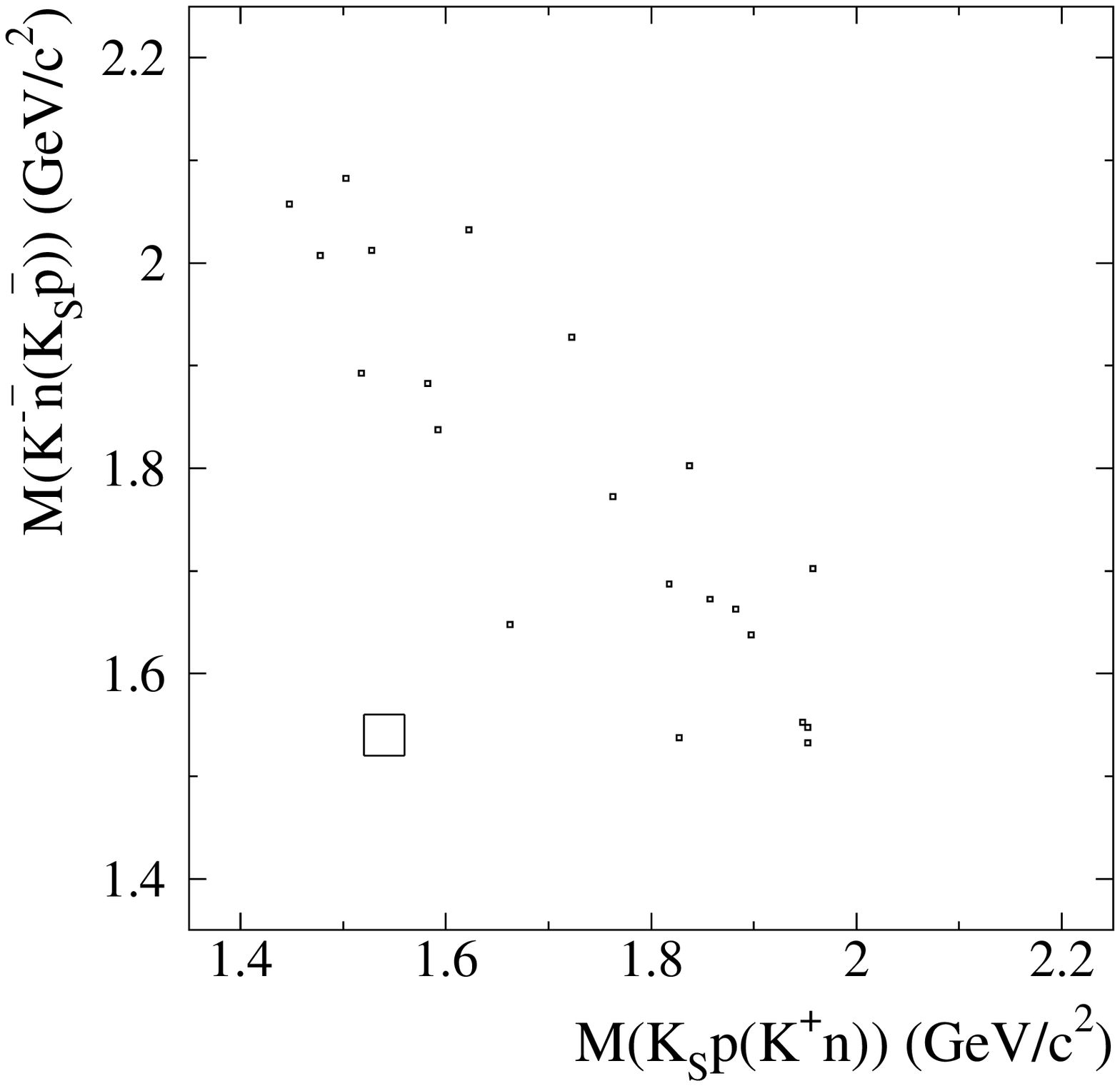}
\vspace*{5pt}
\caption{Scatter plot of $K^-n~(K^0_S \bar p)$  versus $K^0_S p~(K^+ n)$
for
$\psi(2S)\to K^0_SpK^-\bar n$ +
$K^0_S\bar p
K^+n$ modes.}
\end{center}
\end{figure}

Another possibility is that the $\psi(2S)$ decays to only one $\Theta$ or
$\bar\Theta$ state.  To determine the number of $\Theta(1540)$
events from single $\Theta$ or $\bar\Theta$ production, we count the
number of events within regions of 1.52 - 1.56 GeV$/c^2$, shown by the
arrows in Fig. 3 and set upper limits on the branching fractions
at the 90\% C.L.
$${\cal B}(\psi(2S)\to\Theta K^-\bar n\to K^0_S p
K^-\bar n) < 1.0\times 10^{-5}$$
$${\cal B}(\psi(2S)\to \bar \Theta K^+ n\to K^0_S \bar p
K^+ n) < 2.6\times 10^{-5}$$
$${\cal B}(\psi(2S)\to K^0_Sp\bar\Theta\to K^0_S p
K^-\bar n) < 0.60\times 10^{-5}$$
$${\cal B}(\psi(2S)\to K^0_S\bar p\Theta\to K^0_S \bar p
K^+ n) < 0.70\times 10^{-5}.$$
Backgrounds are not subtracted in the calculation of the upper limits.
%Since the statistics of the
%candidate $\psi(2S) \to K^0_SpK^-\bar n$ and
%$K^0_S\bar p
%K^+n$ events is quite low, the effect of backgrounds on the upper
%limit determination is not included.
The numbers used to determine the upper limits  are summarized in Table I. 

\begin{table}[htbp]
\begin{center}
\begin{tabular}{l|ccr}
\hline
\hline
Decay mode~~&~~~N$_{\mbox{obs}}$~~~&~~~~Efficiency~~~~&~~~~Upper limit~~~~\\
\hline
$\psi(2S)\to\Theta\bar\Theta\to K^0_S p K^-\bar n$&&&\\
~~~~~~~~~~~~~~~~~~~$+ K^0_S \bar p K^+
n$&0&$(2.85\pm0.08)\%$&$0.88\times 10^{-5}$\\
$\psi(2S)\to\Theta K^-\bar n\to K^0_S p
K^-\bar n$&1&$(4.07\pm0.09)\%$&$1.0\times
10^{-5}$\\
$\psi(2S)\to\bar\Theta K^+ n\to K^0_S\bar p
K^+ n$&4&$(3.17\pm0.08)\%$&$2.6\times
10^{-5}$\\
$\psi(2S)\to K^0_S p\bar\Theta \to K^0_S p
K^-\bar n$&0&$(3.99\pm0.09)\%$&$0.60\times
10^{-5}$\\
$\psi(2S)\to K^0_S\bar p\Theta\to K^0_S\bar p
K^+ n$&0&$(3.42\pm0.08)\%$&$0.70\times
10^{-5}$\\
\hline\hline

\end{tabular}
\caption{Summary of numbers used in the determination of upper limits for
the $\psi(2S)$ data.}
\end{center}
\end{table}

For the decays of $J/\psi\to K^0_SpK^-\bar n$ and
$K^0_S\bar p
K^+n$, we just use the same
criteria and analysis method as those used for the $\psi(2S)$ data to study
possible $\Theta(1540)$ production. The scatter plot of
$K^-n~(K^0_S \bar p)$ versus $K^0_S p~(K^+ n)$ is shown in Fig. 5.  
 The individual
mass distributions of $J/\psi$ data are shown in Fig. 6, which contains
25 $K^0_SpK^-\bar n$ events 
and 21 $K^0_S\bar p
K^+n$ events.
There is no  $\Theta(1540)$ signal, and we determine
upper
limits
on the
branching fractions at the 90\% C.L.

\begin{figure}[htbp]
\begin{center}
\epsfxsize=7.750cm\epsffile{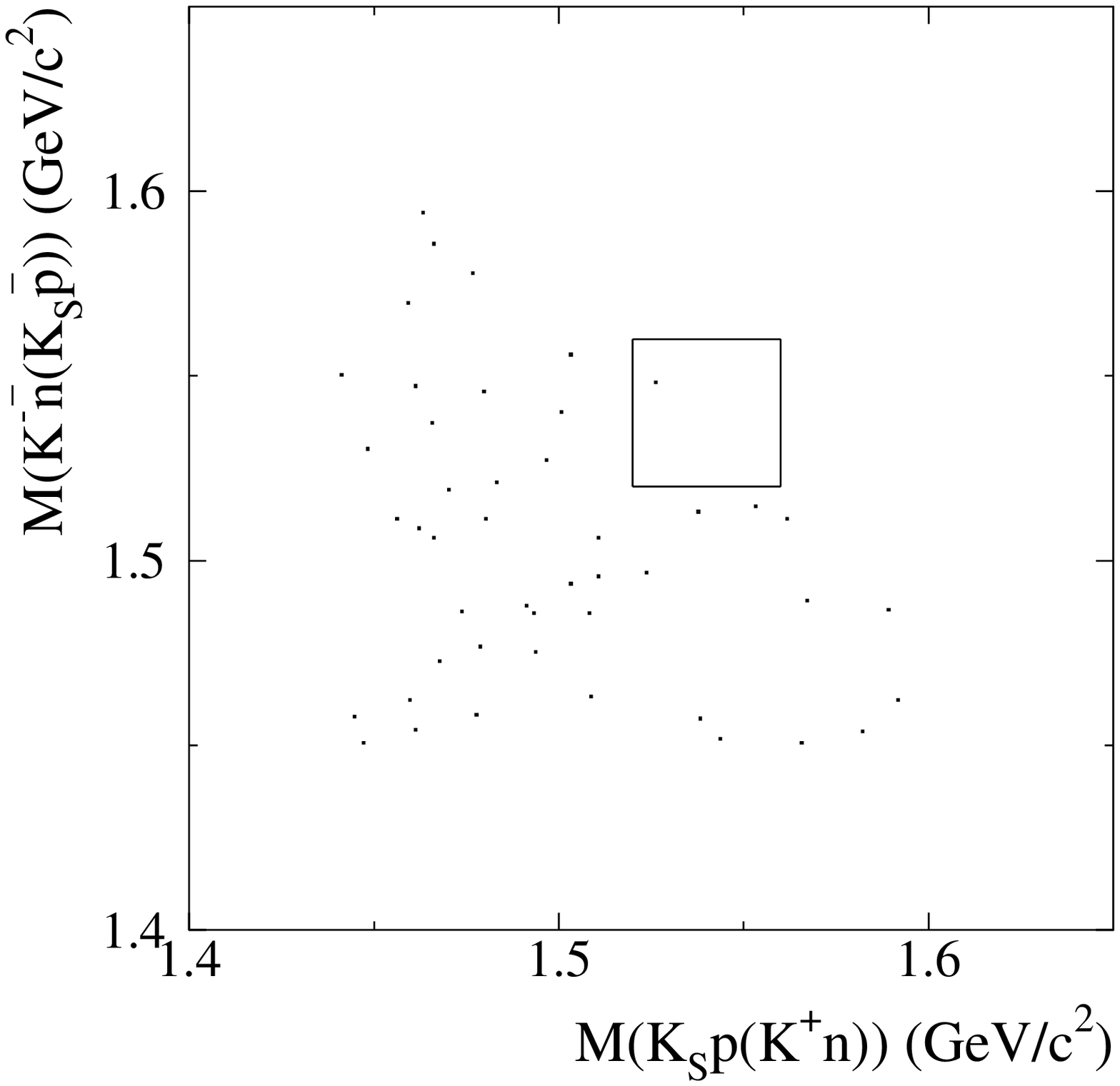}
\vspace*{5pt}
\caption{Scatter plot of $K^-n~(K^0_S \bar p)$ versus $K^0_S p~(K^+ n)$
for
$J/\psi\to K^0_SpK^-\bar n$ +
$K^0_S\bar p
K^+n$ modes.}
\end{center}
\end{figure}

%\vspace*{-25pt}
\begin{figure}[htbp]
\begin{center}
\epsfxsize=12.5cm\epsffile{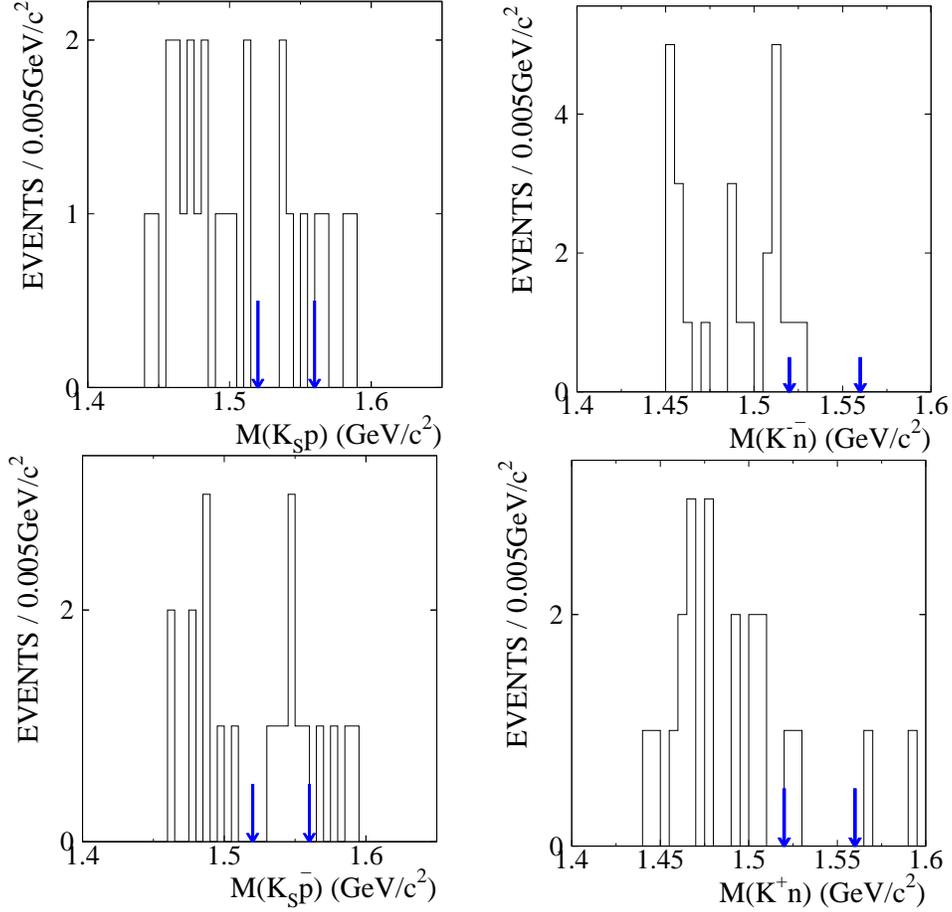}
\label{cpkpa}
\vspace*{5pt}
\caption{Individual mass distributions for the $J/\psi\to K^0_S p K^-\bar
n$ and $K^0_S \bar p K^+n$ modes.
}
\end{center}
\end{figure}

$${\cal
B}(J/\psi\to\Theta\bar\Theta\to K^0_S p K^-\bar n + K^0_S \bar p K^+
n) < 1.1\times 10^{-5}$$
$${\cal B}(J/\psi\to\Theta K^-\bar n\to K^0_S p
K^-\bar n) < 2.1\times 10^{-5}$$
$${\cal B}(J/\psi\to \bar \Theta K^+ n\to K^0_S \bar p
K^+ n) < 5.6\times 10^{-5}$$
$${\cal B}(J/\psi\to K^0_Sp\bar\Theta\to K^0_S p
K^-\bar n) < 1.1\times 10^{-5}$$
$${\cal B}(J/\psi\to K^0_S\bar p\Theta\to K^0_S \bar p
K^+ n) < 1.6\times 10^{-5}.$$\\
These results along with the numbers used to determine them are summarized  in
Table II.

\begin{table}[htbp]
\begin{center}
\hspace*{-2pt}
\begin{tabular}{l|ccr}
\hline
\hline
Decay mode~~&~~~N$_{\mbox{obs}}$~~~&~~~~Efficiency~~~~&~~~~Upper limit~~~~\\
\hline
$J/\psi\to\Theta\bar\Theta\to K^0_S p K^-\bar n$&&&\\
~~~~~~~~~~~~~~~~~$+ K^0_S \bar p K^+
n$&1&~~$(0.88\pm0.04)\%$&$1.1\times 10^{-5}$\\
$J/\psi\to\Theta K^-\bar n\to K^0_S p
K^-\bar n$&4&~~$(0.96\pm0.04)\%$&$2.1\times
10^{-5}$\\
$J/\psi\to\bar\Theta K^+ n\to K^0_S\bar p
K^+ n$&8&~~$(0.59\pm0.03)\%$&$5.6\times
10^{-5}$\\
$J/\psi\to K^0_S p\bar\Theta \to K^0_S p
K^-\bar n$&2&~~$(1.19\pm0.05)\%$&$1.1\times
10^{-5}$\\
$J/\psi\to K^0_S\bar p\Theta\to K^0_S\bar p
K^+ n$&2&~~$(0.86\pm0.04)\%$&$1.6\times
10^{-5}$\\
\hline\hline

\end{tabular}
\caption{Summary of numbers used in the determination of upper limits for
the $J/\psi$ data.}
\end{center}
\end{table}

\section{SUMMARY}

In this work, we studied
$\psi(2S)$ and  $J/\psi$ hadronic decays to $K^0_SpK^-\bar n$ and
$K^0_S\bar p
K^+n$ using 14M $\psi(2S)$ and 58M $J/\psi$ events. No $\Theta(1540)$
signal is observed.
 We set
upper limits for ${\cal
B}(\psi(2S)\to\Theta\bar\Theta\to K^0_S p K^-\bar n + K^0_S \bar p K^+
n) < 0.84\times 10^{-5}$, ${\cal
B}(J/\psi\to\Theta\bar\Theta\to K^0_S p K^-\bar n + K^0_S \bar p K^+
n) < 1.1\times 10^{-5}$ at 90\% confidence level (C.L.).
For the case of single
$\Theta(1540)$ production, the upper
limits determined are also on the order of $10^{-5}$ for
both $\psi(2S)$ and $J/\psi$ data.

\section{Acknowledgements}

   The BES collaboration thanks the staff of BEPC for their hard
   efforts.
This work is supported in part by the National Natural Science
   Foundation
of China under contracts Nos. 19991480,10225524,10225525, the Chinese
   Academy
of Sciences under contract No. KJ 95T-03, the 100 Talents Program of
   CAS
under Contract Nos. U-11, U-24, U-25, and the Knowledge Innovation
   Project of
CAS under Contract Nos. U-602, U-34 (IHEP); by the National Natural
   Science
Foundation of China under Contract No.10175060 (USTC); and by the
   Department
of Energy under Contract No.
DE-FG03-94ER40833 (U Hawaii).

%\newpage %Just because of unusual number of tables stacked at end
%\bibliography{gkk}% Produces the bibliography via BibTeX.

\begin{thebibliography}{99}
\bibitem {WIDTH} We
adopt the approximate width
value determined by recent experiments, which corresponding to $\sim
3\sigma$
in the resolution, 6-7 MeV$/c^2$, at the mass 1.54 GeV$/c^2$.
\end{thebibliography}

\end{document}